\documentstyle[psfig]{mn}

\def\peri{$\cal{T}$}

\begin{document}

\title[The 1997 periastron passage of PSR B1259$-$63]
      {The 1997 periastron passage of the \\ binary pulsar PSR B1259$-$63}

\author[S. Johnston et al.]{S.~Johnston$^1$, N.~Wex$^{2,1}$,
L.~Nicastro$^3$, R.~N.~Manchester$^4$ \& A.~G.~Lyne$^5$\\
$^1$Research Centre for Theoretical Astrophysics, University of Sydney, 
NSW 2006, Australia \\
$^2$Max-Planck-Institut f\"ur Radioastronomie, Auf dem H\"ugel 69, 
D-53121 Bonn, Germany \\
$^3$Istituto di Fisica Cosmica con Applicazioni all'Informatica, CNR,
 Via U. La Malfa 153, I-90146 Palermo, Italy\\
$^4$Australia Telescope National Facility, CSIRO, PO Box 76, Epping, 
NSW 2121, Australia \\
$^5$University of Manchester, Jodrell Bank Observatory, Macclesfield, 
Cheshire SK11 9DL, UK}

\maketitle

\begin{abstract} 
We report here on multifrequency radio observations of 
the pulsed emission from PSR B1259--63 around the time of 
the closest approach to its B2e companion star.
There was a general increase in the pulsar's dispersion measure
and scatter broadening, and a decrease in the flux density towards
periastron although changes in these
parameters were seen on timescales as short as minutes.
The pulsed emission disappeared 16 days prior to periastron
and remained undetectable until 16 days after periastron.

The observations are used to determine the parameters
of the wind from the Be star. We show that a simple model, in which the wind 
density varies with radius as $r^{-2}$, provides a good fit to the data.
The wind is clumpy with size scales $\leq 10^{10}$ cm, densities of
$\sim$10$^{6}$ cm$^{-3}$ and a velocity of $\sim$2000 km~s$^{-1}$ at a distance 
of 20 - 50 stellar radii. We find a correlation between dispersion
measure variations and the pulse scattering times, suggesting that
the same electrons are responsible for both effects.
\end{abstract}


\section{Introduction}
The pulsar PSR B1259$-$63 is a member of a unique binary system. Discovered
using the Parkes telescope in a survey of the Galactic plane at 1.5 GHz
\cite{jlm+92}, it was shown by Johnston et al.~(1992b)\nocite{jml+92}
to be in a highly eccentric 3.4-yr orbit with a 10th-magnitude Be star, SS 2883.

Extensive observations of this system have been made, particularly near
the previous periastron passage on 1994 January 9. These include observations
at radio wavelengths \cite{jml+96,mjsn98,wjm+98,jmmc99}, optical \cite{jml+94},
in the X--ray \cite{crj94,ktn+95,hnt+96} and
in $\gamma$-rays \cite{gtp+95,hckn+99}.
These observations have led to a better understanding of the physical
conditions in the wind and disc of the Be star
\cite{tak94,kc94,mjm95,ta97,jmmc99,bmjs99}.

The understanding of the system is thus as follows:
SS 2883 is a (main-sequence) B2e star with mass $\sim$10 M$_{\odot}$ and
radius $\sim$6 R$_{\odot}$. Its characteristic emission disc extends
to at least 20 R$_*$ in the optical, similar to the distance between
the pulsar and Be star at periastron. It is likely that the disc is
highly tilted with respect to the orbital plane; the orbital plane is itself
at 36$\degr$ with respect to the line of sight. Due to the misalignment
of the emission disc with the orbital plane, the pulsar goes behind
the disc (with respect to the observer) near \peri\ --16 and emerges from
the front of the disc near \peri\ +14. (Here, \peri\ is the
epoch of periastron). The disc material is extremely dense near the
stellar surface but falls off rapidly as a function of distance.
The polar wind, by contrast is $\sim$10$^4$ times less dense than the
disc material at the stellar surface and has an $r^{-2}$ density profile.

The radio observations of the pulsed emission
from the 1994 periastron were rather crude in the 
sense that frequency agility was limited, with frequency switching requiring
manual receiver changes and hence only possible
on a $\sim$1 day timescale. Furthermore, only relatively few observations
were obtained before and after the eclipse of the pulsar. In 1994 there
was no real-time pulse folding software which limited our best
observing strategy in that the off-line data reduction significantly lagged
behind the observing. However, these observations showed dispersion measure
(DM) and scintillation parameter changes on timescales of at least 
as short as one day
and demonstrated that these changes were highly asymmetric with 
respect to periastron \cite{jml+96}.

In this paper we report on observations of the pulsed radio emission made 
around the latest periastron passage of the pulsar on 1997 May 29. Section 2 
describes the observations. In Section 3 we present the results and 
in Section 4 discuss their implications.

\section{Observations}
In 1996, a major upgrade was made to the focus cabin of the Parkes
radio-telescope which allowed several receivers to be
placed on a movable platform. This enables the choice of receiver to be
made from the control room and allows for frequency changes in
only a few minutes.

We observed PSR B1259--63 regularly with the 64-m Parkes radio telescope 
at frequencies of 1.4, 4.8 and 8.4 GHz \cite{wjm+98}. The observations
were intensified around the periastron passage of the pulsar on 1997 May 29.
At each frequency the receiver consists of
a dual-channel, cryogenically cooled system sensitive to two orthogonal
linear polarizations.

Data were recorded simultaneously using two recording systems.
The Caltech Fast Pulsar Timing Machine (FPTM) is a correlator capable 
of fast sampling \cite{nav94}. The signals from the two
polarizations are down-converted to intermediate frequency and
2-bit digitised by the FPTM which computes
correlation functions (CFs) with 128 lags over a 128 MHz 
bandwidth. The system can be used either to record full Stokes parameters for
one 128 MHz bandwidth or total intensity for two independent
bandwidths. For the higher frequencies we always recorded the data
in full Stokes mode, with the band centered at 4800.25 and 8400.25 MHz.
In the 20cm band, we either recorded the full Stokes parameters at 1351.0 MHz
or two independent bandwidths centered at 1274.75 and 1396.0 MHz.
The CFs are then hardware integrated at the apparent pulsar period 
($\sim$47 ms) and the data recorded every 60~s. A graphical output
displays the pulse profile every integration cycle. Thus, changes in 
observing strategy could be made in real-time.
  
Simultaneously, we also recorded data using a filterbank
detection system in an identical fashion to that described elsewhere
(e.g. Johnston et al. 1996\nocite{jml+96}).
The signals from the receivers were down-converted to
an intermediate frequency and then passed into a filterbank which
consisted of 64 channels, each 5 MHz wide for a total of 320 MHz.
Full Stokes parameters were recorded every 0.6 ms and written to magnetic tape 
for off-line analysis. The centre frequencies were 1374, 4748 and 8348 MHz.

For flux density calibration, observations were made of Hydra A at
1.4, 4.8 and 8.4 GHz once every few days at the telescope. These
measurements yielded system flux densities of $\sim$28,
$\sim$100 and $\sim$100 Jy at the three frequencies.
Before every observation of the pulsar, a pulsed
noise source, injected at 45 degrees to the signal probes, was recorded.
This allows for calibration of the system gains and phases.

\subsection{Derivation of the Dispersion Measure}
Over a period of a few hours
we obtained observations at 5 different frequencies:
1.2, 1.4, 1.5, 4.8 and 8.4 GHz. The three frequencies in the 20-cm band were
obtained simultaneously. The observation would then be followed by one
at either 4.8 and 8.4 GHz before again observing the 20-cm band. Typical
observations were 1-hr in duration. Pulse time-of-arrivals were
obtained in the standard manner (e.g. Wex et al. 1998\nocite{wjm+98}).
The DM was obtained
by assuming the DM contribution from the 1.5 kpc path through the
interstellar medium  is 147.6 cm$^{-3}$pc and does not change
with time. Then, the extra DM contribution from the
wind of the Be star could be computed
by the relative differences in the residuals between the different
frequencies. Having closely spaced frequencies in the 20-cm band
allows any ambiguities to be resolved (i.e. where the $\Delta$DM could
impose more than 1 phase turn on the residual). Typical errors using
this method are small ($\sim$0.1 cm$^{-3}$pc), much smaller than
those obtained simply from fitting the delays in frequency-phase
space over a single observing band.

\subsection{Derivation of the Scattering Time}
Scatter-broadening of a pulse can be approximated by convolving the
intrinsic pulse shape with a truncated exponential function.
The situation for PSR B1259--63 is somewhat complicated
by the presence of two strong pulses, only 20 ms apart in time \cite{mj95}.
For large scattering times, the scatter tail of the first pulse overlaps
the second pulse and this must be taken into account in the derivation
of the scattering time. We used the Levenberg-Marquart method (LMM) to
perform a least-squares fit to the observed profile given a `standard'
(unscattered) profile. The fit returns the 4 free parameters : amplitude,
baseline and decay time of the exponential and the offset in time
between the profiles. Generally, for scattering times less than 7 ms,
the fits are good. Above this, however, the pulses tend to be weak,
and the extremely scattered and weak
pulses make the fitting less robust. This is reflected in much larger errors
bars on the fitted parameters.

\section{Results}
The broad overview of the observational data is as follows.
A change in the DM over and above that from the ISM
was first observed at \peri\ --48. Over a period of $\sim$20 days, both
the DM and the pulse scattering showed an overall increase, and strong
hourly variations. The pulsed emission was last detected on \peri\ --17.
Observations made on \peri\ --15 showed no pulsed emission at any frequency,
a result confirmed with independent observations on the
Australian Telescope Compact Array \cite{jmmc99}. 
The pulses were re-detected at Parkes on \peri\ +16,
following a non-detection two days earlier. Post-periastron, no change
in DM or pulse-scattering could be measured, although significant changes
in rotation measure and polarization properties were recorded.
Significant flux variability was seen both prior to and 
following the eclipse. The general trend though was towards lower fluxes
in the few days before and after the eclipse.

Figure 1 shows the DM changes and the scattering time, $\tau$,
as a function of epoch prior to periastron. Figure 2 shows a shorter
time span of only 8 days prior to the eclipse.
\begin{figure}
\psfig{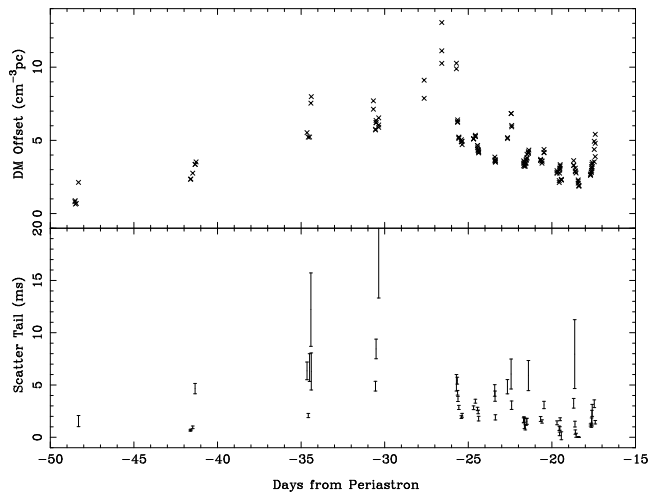}
\caption{Dispersion Measure changes and derived scattering times as a function
of time prior to periastron. The errors on the DM are of order 0.1 cm$^{-3}$pc
and are too small to be seen on the plot.}
\end{figure}
\begin{figure}
\psfig{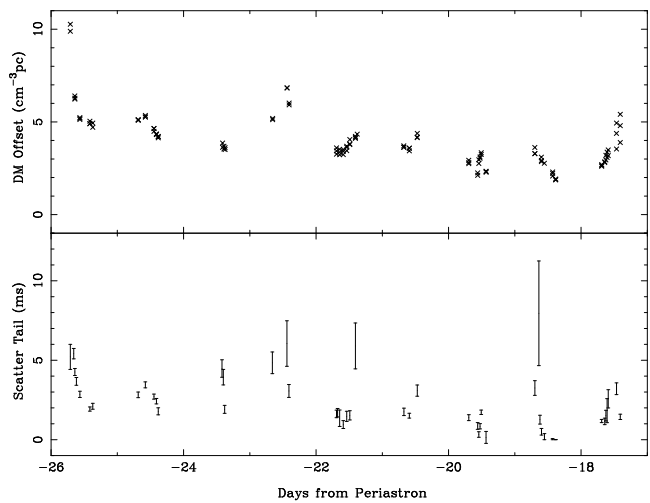}
\caption{As Fig. 1, showing a period of only 8 days just prior to the
eclipse of the pulsar.}
\end{figure}

The figures show a wealth of detail. Points to note are that the DM changes
on short timescales ($\sim$1 hr or less) by $\sim$1 cm$^{-3}$pc.
There is a slow increase in the DM from \peri\ --50 until --27.
The DM then slowly drops, before rising steeply again at \peri\ --17.
The pulsar was undetected following these observations until \peri\ +16.
Both the daily variations and the drop in DM towards \peri\ --20 are
evidence for the clumpy nature of the wind.

\begin{figure}
\label{flux}
\psfig{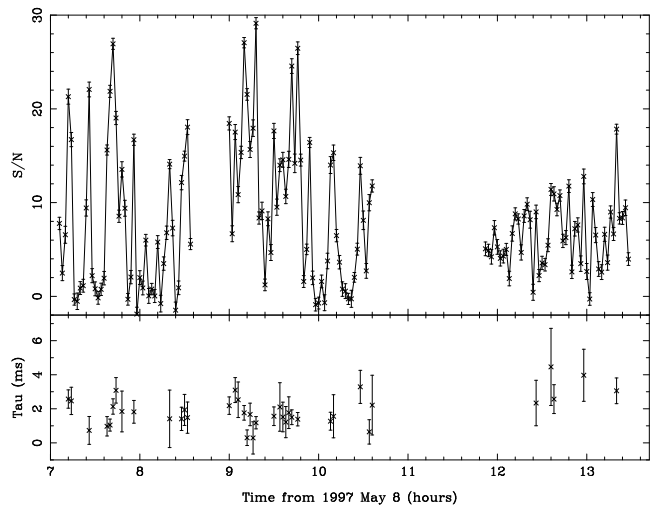}
\caption{Top panel : Flux density of PSR B1259--63 on 1997 May 8.
The observing frequency is 1.4 GHz and the fluctuations are broadband.
Lower panel : Scattering time in ms for those integrations with large
enough s/n to obtain reliable estimates.} 
\end{figure}

Flux density variations in PSR B1259--63 occur on a number of timescales.
The diffractive scintillation timescale at GHz frequencies close
to periastron is expected
to be very short ($\ll 1$ second) and hence unobservable. The refractive
timescale is probably of the order of minutes. One also expects
changes in the flux density due to free-free absorption as the pulsar
moves behind clumps in the Be star wind. In Figure \ref{flux}, we show
flux variations as a function of time starting on 1997 May 8.3
(\peri\ --21).
Each point is an integration over 2 minutes of data.
These variations are broad band (at least across the 300 MHz
observing bandwidth) at 1.4 GHz. At higher frequencies, the pulsar
is too weak to allow reliable flux densities to be obtained in a few minutes.

This extreme flickering is only observed on \peri\ --21.
Observations made the previous day and on the following three days show
only slow changes in flux density over hours and a much lower
modulation index, consistent with the refractive scintillation timescale.
\peri\ --21 corresponds exactly to the first 
enhancement in the flux density seen in the unpulsed source \cite{jmmc99}.
This early increase in the flux density subsequently disappeared and
does not fit well with the model of the unpulsed source \cite{bmjs99},
although Johnston et al. (1999) have argued that it is likely to be
the result of the first interaction between the pulsar wind and 
the stellar disc coupled with optical depth effects.

\section{Discussion}
At \peri\ --20, the pulsar is $\sim$50 stellar radii from the Be star.
Assuming a wind density at the stellar surface of $7\times 10^8$cm$^{-3}$
which varies with radius as $r^{-2}$, the line-of-sight density to the pulsar
should be $\sim$3$\times 10^5$cm$^{-3}$. However, one cannot expect
this wind to be homogeneous and there are likely to be significant density
variations because of turbulence.

We can use the equation derived by McClure-Griffiths et al. (1998)
\nocite{mjsn98} to relate the electron density in the scattering material,
$\delta n_e^2$, to the scattering parameters.
\begin{equation}
\delta n_e^2 = 4.0 \,\,\,
\Delta \nu^{-5/6}_{\rm d,MHz} \,\,\,
D^{-5/6}_{\rm kpc} \,\,\,\,\, \nu^{11/3}_{\rm GHz} \,\,\,
\frac {l_{o,{\rm pc}}^{2/3}} {\Delta L_{\rm pc}} \,\,\,
\left[ \frac {(1+x)^2}{x}\right] ^{5/6} \,\,\,{\rm cm}^{-6}
\end{equation}
Here, $\Delta \nu$ is the scintillation bandwidth, $D$ the distance
to the pulsar and $\nu$ the observing frequency. $l_o$ is the outer
scale length, $\Delta L$ is the thickness of the scattering screen and $x$ is
the ratio of the observer-screen distance and the pulsar-screen separation.
We simplify this by using $D=1.5$ kpc, $\nu=1.4$ GHz and assuming
$l_o \approx \Delta L$, $x \gg 1$ to obtain
\begin{equation}
\delta n_e^2 = 9.8 \,\,\,\,\,
\Delta \nu^{-5/6}_{\rm d,MHz} \,\,\,\,\, x^{5/6} \,\,\,{\rm cm}^{-6}
\end{equation}
Converting the scintillation bandwidth, $\Delta \nu$ to the scattering
time, $\tau$, in milliseconds via $2\pi\tau\Delta\nu=1$ yields
\begin{equation}
\delta n_e^2 = 1.43\times 10^4 \,\,\,\,\,
\tau^{5/6} \,\,\,\,\, x^{5/6} \,\,\,{\rm cm}^{-6}
\end{equation}
Assuming a distance to the screen, $r_s$,  of $\sim 8.4\times 10^{12}$cm,
(i.e $x=5.5\times 10^8$), a scattering time of 3 ms gives
$\delta n_e^2$ of $7\times 10^{11}$cm$^{-6}$ and hence
$n_e$ of $\sim 8.3\times 10^5$cm$^{-3}$ assuming a fully modulated wind.
This is remarkably close to
the expected value obtained from a simple scaling of the electron density
in the wind and shows that the scattering is caused
by the global properties of the fully modulated wind.
The size of the scattering disc, given by $(2\tau c r_s)^{1/2}$ is rather small,
only $3.9\times 10^{10}$ cm.

These values are consistent with the DM measurements. A mean wind
density of $\sim 8.3\times 10^5$cm$^{-3}$ along a pathlength of
roughly $2r_s$ yields a $\Delta$DM of 4.5 cm$^{-3}$pc, close to the
observed values. We also note that to double the observed scattering tail
requires only a change in $n_{e}$ of $\sim$30 per cent and that this
change affects the DM at the $\sim$1 cm$^{-3}$pc level. Such changes
are seen on roughly a 2-hr timescale.

\begin{figure}
\label{dmvtau}
\psfig{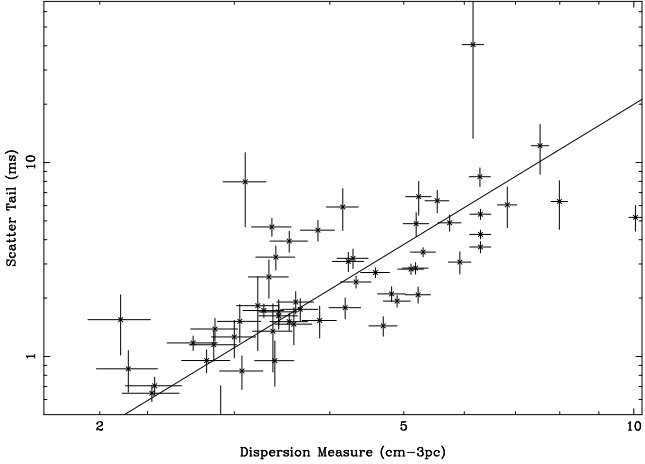}
\caption{Scattering time versus change in DM at a given epoch. The
straight line shows $\tau \propto {\rm DM}^{12/5}$.}
\end{figure}

From equation 3 above it is evident that 
$\tau\propto n_e^{12/5}$ whereas DM $\propto n_e$.  Providing the same
electrons are responsible for the scattering and the dispersion
we should observe that $\tau \propto {\rm DM}^{12/5}$.
Figure 2 shows the expected correlation between dispersion measure and
scattering time in that they both rise and fall together.
In Figure \ref{dmvtau} we show a plot of scattering time versus $\Delta$DM.
The scatter in the diagram leads to a large uncertainty in any formal
straight line fit. For illustrative purposes, we show 
a line with slope 12/5 as derived above, which is in satisfactory
agreement with the data.

Consider the free-free optical depth along the line of sight, given by
\begin{equation}
\tau_{\rm ff} = 8.2\times 10^{-2} \,\,\,\,\, T^{-1.35} \,\,\,\,\, \nu^{-2.1}
\int_{0}^{d} n_{e}^{2} \,\,\,\,\, {\rm dl}
\end{equation}
where $T$ is the temperature in Kelvin, $\nu$ the observing frequency
in GHz, $d$ the distance through the absorbing material in pc and $n_e$ the free
electron density within the clump.  For an electron
temperature in the wind of $\sim 10^{4}$ K \cite{wat86}, the wind parameters
described above yield $\tau_{\rm ff}\sim 0.6$ at 1.4 GHz, fully
consistent with the lower flux densities recorded at these epochs.
Flux density variations are seen both on timescales of a few minutes and
on a roughly similar timescale to the DM and scattering variations.
The longer timescale is probably due to variations in the free-free
optical depth, either because of temperature variations or small variations
in the electron density. The short timescale variations generally have
a low modulation index and are likely to be refractive scintillation.
Given a timescale of $\sim$200 s for the variations and knowledge of
the size of the scattering disc ($3.9\times 10^{10}$cm), the dominant
velocity in the system is then $\sim$2000 kms$^{-1}$. Such a velocity
is consistent with UV observations of Be star winds in general.

In summary, we have shown that a plausible model for 
the wind of the Be star can account for the dispersion measure
and scattering data before periastron and is also consistent with
the variability timescale and the flux density variations. We stress
that we are discussing the spherical, high-velocity, low-density
wind from the star. The pulsar encounters the circumstellar disc material
at \peri\ --18, the density of the disc is so large that the pulsed flux
becomes eclipsed shortly thereafter through a combination of free-free 
absorption and extreme pulse broadening and remains so until the pulsar 
re-emerges from the disc at \peri\ +16. Post-periastron, the pulsar 
lies between the observer and the Be star. The path length through
the stellar wind is much less than pre-periastron and the DM and
scattering variations are hence much smaller.

\section*{Acknowledgments}
We thank the Parkes staff, especially H. Fagg and M. McColl, for 
their considerable help throughout the observations.
We thank M. Britton for useful discussions.
The Australia Telescope is funded by the Commonwealth of Australia
for operation as a National Facility managed by the CSIRO.

\bibliographystyle{mn}
\bibliography{modrefs,psrrefs}

\end{document}